\documentclass[
twocolumn,
hf,
]{ceurart}

\usepackage{listings}
\usepackage{color}

\begin{document}

\pagestyle{plain}
\pagenumbering{arabic}

\copyrightyear{2023}
\copyrightclause{Copyright for this paper by its authors.
  Use permitted under Creative Commons License Attribution 4.0
  International (CC BY 4.0).}

\conference{IJCAI 2023 Workshop on Deepfake Audio Detection and Analysis (DADA 2023), August 19, 2023, Macao, S.A.R}

\title{TranssionADD: A multi-frame reinforcement based sequence tagging model for audio deepfake detection}


\author[]{Jie Liu}[%
email=jie.liu5@transsion.com
]
\fnmark[1]

\author[]{Zhiba Su}[%
email=zhiba.su@transsion.com
]
\fnmark[1]

\author[]{Hui Huang}[%
email=hui.huang6@transsion.com
]

\author[]{Caiyan Wan}[%
email=caiyan.wan@transsion.com
]

\author[]{Quanxiu Wang}[%
email=quanxiu.wang@transsion.com
]

\author[]{Jiangli Hong}[%
email=jiangli.hong@transsion.com
]

\author[]{Benlai Tang}[%
email=benlai.tang@transsion.com,
]
\cormark[1]

\author[]{Fengjie Zhu}[%
email=fengjie.zhu@transsion.com
]

\address[]{Department of AI Technology, Transsion}

\cortext[1]{Corresponding author.}
\fntext[1]{These authors contributed equally.}


\begin{abstract}
Thanks to recent advancements in end-to-end speech modeling technology, it has become increasingly feasible to imitate and clone a user’s voice. This leads to a significant challenge in differentiating between authentic and fabricated audio segments. To address the issue of user voice abuse and misuse, the second Audio Deepfake Detection Challenge (ADD 2023) aims to detect and analyze deepfake speech utterances. Specifically, Track 2, named the Manipulation Region Location (RL), aims to pinpoint the location of manipulated regions in audio, which can be present in both real and generated audio segments. 
We propose our novel TranssionADD system as a solution to the challenging problem of model robustness and audio segment outliers in the trace competition. 
Our system provides three unique contributions: 1) we adapt sequence tagging task for audio deepfake detection; 2) we improve model generalization by various data augmentation techniques; 3) we incorporate multi-frame detection (MFD) module to overcome limited representation provided by a single frame and use isolated-frame penalty (IFP) loss to handle outliers in segments. Our best submission achieved 2nd place in Track 2, demonstrating the effectiveness and robustness of our proposed system.
\end{abstract}

\maketitle

\section{Introduction}

In recent years, technologies such as speech synthesis \cite{wang2017tacotron, ren2020fastspeech, tan2022naturalspeech, shen2023naturalspeech, luong2020nautilus, ma2020fpets} and voice conversion \cite{casanova2022yourtts, li2023freevc, wang2021towards} have developed rapidly and have reached a point where they can deceive people into thinking the audio is real. Current researches have gone deep into anti-spoofing model \cite{wu2020defense, wang2021comparative} and defense techniques for automatic speaker verification (ASV) model \cite{peng2021pairing}. The second Audio Deepfake Detection Challenge (ADD 2023) \cite{deepfake} aims to encourage researchers to boost technologies of detection and analysis of deepfake speech utterances. The objective of track 2 in ADD 2023 is not limited to identifying fake or real audio, but it also involves identifying the location of the fabricated audio clips, which presents a more complex challenge for us.

Most detection methods focus on predicting binary labels of "real" and "fake". Zeng et al. \cite{zeng4051713deletion} use a universal background model (UBM) to extract Electric Network Frequency (ENF) fluctuation super vector and then build a deep neural network to train and classify the ENF fluctuation super vector. Their further work \cite{zeng2022} uses network constructed by CNN \cite{gu2018recent} and LSTM \cite{6795963} to extract deep spatial and temporal features respectively. A technique \cite{wang2022} that integrates shallow and deep features is proposed to help the model learn features at various levels and recognize inconsistencies resulting from tampering operations. Besides ENF, the Mel-frequency cepstral coefficients (MFCCs) \cite{winursito2018improvement} are used to attain valuable representation to distinguish deepfake \cite{hamza2022}. These works are mainly concentrated on traditional audio processing techniques, which are hard to obtain and require expert knowledge diving into signal processing. Moreover, shallow networks currently employed have poor performance in vivid speech synthesis systems. RawNet2 \cite{tak2021end} directly uses raw waveform with 64000 samples as input and proposes a convolutional neural network architecture for speaker verification. However, as RawNet2 is specifically designed for classification tasks, it is unable to handle outliers in time series data and does not incorporate the fusion of contextual information.

Moreover, to locate specific areas of manipulated regions, several methods have been proposed. In \cite{ustubioglu2023}, Scale-Invariant Feature Transform (SIFT) keypoints, which are obtained from each RGB color channel of mel-spectrograms image, are matched via feature vector to locate forged regions. The center of the forged regions is used to mark the forged segments according to correlation features and euclidean distances. Alternatively, Yang et al. \cite{yang2023} convert the problem of detecting and locating duplicates into a procedure similar to locality-sensitive hashing. Neither of these two works is suitable for situations where the forged region is extremely similar to its neighbor, such as a segment forged by noise addition only. Wu et al. \cite{wu2022partially} introduce a location model similar to Question-Answering strategy, which is to make the model answer "where are the start and end points" of anomalous clips. But it makes the problem more ambiguous because apparently it is not a natural language understanding task.

To solve these problems, we propose TranssionADD system, which specifically addresses poor generalization and outlier problems of the location model. Firstly, we innovatively convert the task into a sequence tagging problem, predicting the label of each frame and merging frames with the same label to locate audio segments. Secondly, we use RCNN-BLSTM to extract spatial and temporal features frame by frame. To avoid the network's poor learning of contextual information, we use multi-frame detection (MFD) module to compress multi-frame information. Then we address data insufficiency by applying data augmentation including voice conversion, pitch shift, etc. Finally, to solve outliers from isolated frames, we introduce a penalty strategy called isolated-frame penalty (IFP) to constrain this situation. Our submission won second place in the ADD 2023. More explicitly, our system mainly contributes to the following parts:


\begin{itemize}
    \item We design a novel location model that draws inspiration from the sequence tagging task, specifically the Name Entity Recognition (NER) task.
    \item We utilize data augmentation method both before training and during training.
    \item We apply MFD module to improve robustness and add IFP loss to deal with outliers.
    \item Experiments show our proposed method performs better than the baseline.
\end{itemize}

\section{Method}

In this section, we introduce details and rationales of our proposed method, which mainly consists of four parts: 
1) RCNN-BLSTM backbone; 2) data augmentation; 3) multi-frame detection module; and 4) isolated-frame penalty loss.

\begin{figure*}[!htbp]
\centering
\includegraphics[width=1.0\linewidth, height=0.6\linewidth, trim = 10 10 10 10, clip]{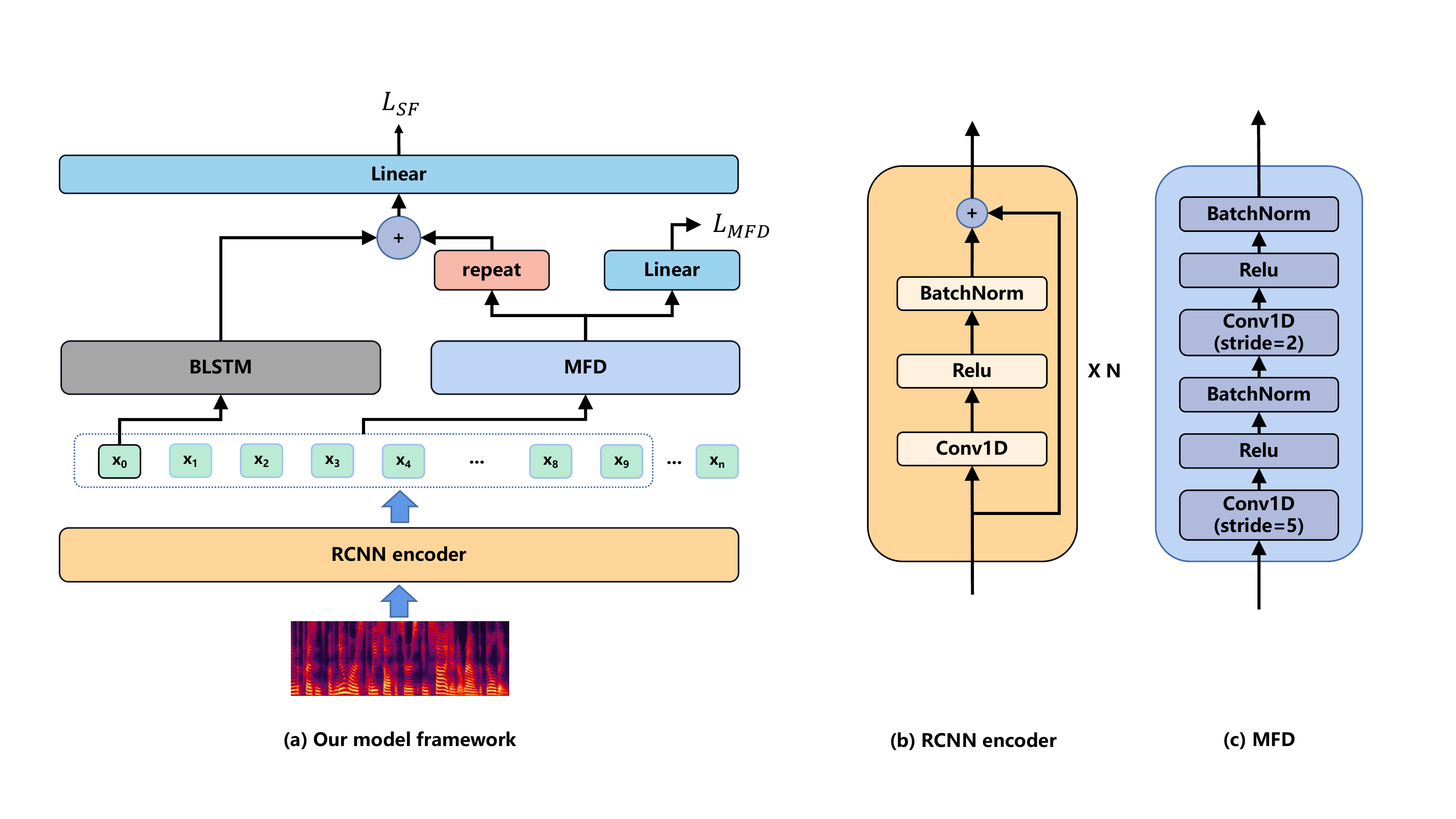}
\caption{Overview of our proposed system. Subfigure (a) illustrates the overall structure of our model. Subfigure (b) is the RCNN encoder. Subfigure (c) shows the multi-frame detection module (MFD).} 
\label{fig:model}
\end{figure*}

\subsection{RCNN-BLSTM Backbone}
 As shown in Figure \ref{fig:model}, we use mel-spectrograms extracted from audio as input representations, 
and then we build a basic model based on residual convolutional neural network (RCNN) and bi-directional long short-term memory (BLSTM). Some of our structures refer to the design of RawNet2 \cite{tak2021end} and reference encoder \cite{skerryryan2018referenceencoder}.
Mel-spectrograms are passed through 7 layers of 1D residual convolution with filter size and kernel size set to be 256 and 3 respectively. After each convolutional layer, batch normalization and Relu activation are applied.

Next, the output of final convolutional layer is fed into a single bidirectional LSTM layer with 256 units (128 in each direction) to generate classification representations.
At the end of the model, we use a 
linear layer to output the classification probabilities
of each frame of audio.

\subsection{Data Augmentation}

In this study, there are two major difficulties: 1) the data is very limited and imbalanced; 2) the training and test sets have huge disparities in terms of their domain. The model, therefore, suffers from underperformance. To address this issue, we propose a two-stage data augmentation method.

\textbf{Before-training augmentation.} Both the training set and dev set are insufficient and lack diversity. Therefore, we propose three  data augmentation methods: 
\begin{itemize}
    \item \textbf{Voice conversion.} Use VC \cite{li2023freevc} model trained on given dataset to transform segments of audio and then annotate adjusted segments as "fake".

    \item \textbf{Noise adding.} Add noise from MUSAN \cite{musan2015} corpus to entire audio segments.
    \item \textbf{Insertion.} Insert a real audio clip with different lengths randomly selected from other audio or the audio itself.
\end{itemize}

\textbf{In-training augmentation.} Due to the fixed nature of data augmentation before training, it limits the randomness of the data. Therefore, we propose dynamic data augmentation during training to ensure sufficient diversity and randomness in the training data.
\begin{itemize}
    \item \textbf{Pitch shift.} Randomly adjust the pitch of segments from audio and then label the adjusted segments as "fake".
    \item \textbf{Gaussian noise.} Add Gaussian noise to segments from audio and then label the adjusted segments as "fake".
\end{itemize}

\subsection{Multi-frame Detection Module}
RCNN primarily focuses on single-frame information but lacks a broader context. To further extract more effective contextual information, we draw inspiration from wav2vec2.0 \cite{baevski2020wav2vec} and enhance current single-frame detection with an additional multi-frame detection (MFD) module trained in a supervised manner.
Multiple audio frames are passed through a downsampling network to get their multi-frame feature. The logits of the average label values corresponding to these multiple audio frames are calculated to get an estimated label of this segment. The multi-frame feature is then passed through a classification network using the estimated label as the target.

Precisely, as shown in Figure \ref{fig:model} , we use two convolutional layers to build the downsampling network and one dense layer to form the classification network. The filter size of both convolutional layers is set to 128. The stride values are set to be 5 and 2; the kernel sizes are set to be 7 and 3, respectively. The cross-entropy loss of MFD is added to the training loss.
Finally, we replicate and add the outputs of the MFD module to the outputs of BLSTM for each frame, to enhance the classification prediction accuracy for each frame.

\subsection{Isolated-frame Penalty Loss}

In traditional NER tasks, one entity corresponds to one or several word(s); while in ADD tasks, one fake/real segment corresponds to tens or hundreds of frames. Consequently, when applying regular NER to audio frame-level predictions, a challenge arises wherein the model predicts labels that appear on a significantly higher number of discontinuous frames, which we refer to as outliers. Therefore, in order to constrain the case where the predicted value is different from the surrounding values, we design an additional loss function called isolated-frame penalty (IFP) loss.

Specifically, we devise a penalty calculation method: simultaneously calculating the difference between current frame and its surrounding frames (1 to 3 before and after).
Particularly, for frame $i$, its predicted probability is $\hat{y_i}$, and we 
consider {the previous and following }$s$ frames, $s\in \{1,2,3\}$ . Correspondingly, we can obtain the regularization term $r_i^s$.

\begin{equation}\label{eq:l2}
    r_i^{(s)} = ||\hat{y}_{i}-\frac{\sum_{k=1}^{s} (\hat{y}_{i-k}+\hat{y}_{i+k})}{2s} ||    
\end{equation}

Finally, we calculate the regularity constraint of total N frames and divide the sum by $3$ as the final regularity constraint term $R$. The formula is as follows:
\begin{equation}\label{eq:lr}
  R=\sum_{i=1}^N(r_i^{(1)}+r_i^{(2)}+r_i^{(3)})/3
\end{equation}


 Then, our total loss involves losses of multiple modules, and its expression is as follows:
\begin{equation}\label{eq:l6}
  L=L_{SF}+L_{MFD}+\alpha*R
\end{equation}


Where $L_{SF}$ and $L_{MFD}$ represent the cross-entropy loss for the single frame and each MFD module, respectively, while $R$ serves as the constraint applied to isolated small fragments. It is important to note that the weight $\alpha$ assigned to the IFP loss is a hyperparameter.
\section{Experiments}

\begin{table*}[[!htbp]]
\caption {\textbf{Scores} of the dev/test dataset and the augmented dev dataset, and the \textbf{iso-rate} of their respective auxiliary metrics. We conduct a series of controlled experiments using modified Rawnet2, RCNN + BLSTM backbone (proposed baseline), and our TranssionADD (RCNN + BLSTM + two-stage data augmentation + MFD module + isolated-frame-penalty) as the experimental control groups. The best in \textbf{bold} in each metric.}
  \label{tab:freq}
  \begin{tabular}{l|ccccc}
    \toprule[1.5pt]
    Model & dev-score↑ & iso-rate↓(\%) & aug-dev-score↑ & aug-iso-rate↓(\%) & test-score↑ \\
    \midrule[1.5pt]
    Rawnet2 (baseline) & 0.9911 & 0.2245 & 0.9502 & 1.4183 & 0.2783 \\
    
    RCNN+BLSTM (proposed baseline) & 0.9926 & 0.1684 & 0.9535 & 1.40 & 0.2842  \\







    \textbf{TranssionADD} & \textbf{0.9957} & \textbf{0.1123} & \textbf{0.9933}  & \textbf{0.6873} & \textbf{0.6249} \\

  \bottomrule[1.5pt]
\end{tabular}
\end{table*}

\begin{table*}[[!htbp]]
\caption  {Performance comparison for ablation studies. (In the table, "-" means to remove the corresponding module)}
  \label{tab:freq1}
  \begin{tabular}{l|ccccc}
    \toprule[1.5pt]
    Model & dev-score↑ & iso-rate↓(\%) & aug-dev-score↑ & aug-iso-rate↓(\%) & test-score↑ \\
    \midrule[1.5pt]
    \textbf{TranssionADD} & 0.9957 & \textbf{0.1123} & \textbf{0.9933}  & \textbf{0.6873} & \textbf{0.6249} \\
    \midrule[1.0pt]
    \quad -IFP & \textbf{0.9959} & 0.4153 & 0.9931 & 1.0865 & 0.6236 \\
    \quad\quad -MFD & 0.9934 & 0.2245 & 0.9885 & 0.8692 & 0.5182 \\
    \quad\quad\quad -InAug & 0.9953 & 0.1347 & 0.9804 & 1.5992 & 0.4126  \\
    \quad\quad\quad\quad -BefAug & 0.9926 & 0.1684 & 0.9535 & 1.40 & 0.2842  \\








  \bottomrule[1.5pt]
\end{tabular}
\end{table*}

\subsection{Datasets}
Our dataset is a publicly available Mandarin corpus provided by organizers of ADD 2023. 
Some samples are all real or all fake, and some are partially fake or partially real.  
\subsection{Training Setup}
\subsubsection{Input representations }
We compare mel-spectrograms and MFCCs, and finally we choose mel-spectrograms as low-level acoustic representation due to the better performance. We use a 16 kHz sampling rate for all experiments. And window size of fast Fourier transform (FFT), hop size, and number of output bins are set to 800, 160, and 80 respectively.

\subsubsection{Data augmentation}
Before training, we augment the training set and dev set using techniques such as VC, noise addition, and insertion. During training, we use a 0.2 random rate to add Gaussian noise with less than 15 dB signal-to-noise ratios (SNR) and change the audio pitch by pitch shift.

\subsubsection{Implementation details}
We first train the model on one NVIDIA 3090 GPU, with 64 batch sizes and 100 epochs. The models are optimized by Adam with a learning rate of 0.0001 and weight decay of $10^{-5}$. The hyper-parameter weight $\alpha$, in the isolated-frame penalty (IFP) loss, is set to be 0.1 as default, based on simple tuning experiments.

\subsubsection{Metrics}
As follow, $A_{sentence}$ is used to measure the model's ability to correctly distinguish between genuine and fake audio, and $F_{1\_segment}$ is used to measure the model's ability to correctly identify fake areas from fake audios. The final score is the weighted sum of $A_{sentence}$ and $F_{1\_segment}$.
\begin{equation}
score=0.3*A_{sentence}+0.7*F_{1\_segment}
\end{equation}

Moreover, we consider audio should consist of long enough consecutive frames with the same label and isolated frames are possible to be outliers. Therefore, relying solely on $A_{sentence}$ and $F_{1\_segment}$ may not adequately capture the occurrence of fragmented and isolated segments in audio frame predictions. To evaluate the efficacy of dealing with outliers, we introduce a simple auxiliary evaluation metric called "iso-rate," as defined in Eq.~\ref{eq:l5}. If a segment is shorter than 6 frames (60ms), we classify it as an isolated segment ($N_{isolated}$), and then compute the proportion of these isolated segments appearing across the entire dataset. A lower iso-rate indicates that the evaluated system has fewer isolated segments and addresses outliers better.

\begin{equation}\label{eq:l5}
  \textit{iso-rate} = \frac{N_{{isolated}}}{N_{{audios}}}
\end{equation}

\subsection{Results}

From Table \ref{tab:freq} above, our system achieves significantly higher scores on the development set than on the test set which may contains more variable and complicated samples. To address this problem, we introduce the augmented development set (aug-dev) as an essential reference to evaluate the model's performance.

\subsubsection{Performance}

Table \ref{tab:freq} compares three systems: 1) a modified open-source implementation, Rawnet2; 2) our proposed baseline, which consists of RCNN and BLSTM; and 3) our best system, TranssionADD. The results show that TranssionADD significantly outperforms the other two systems in terms of both correct detection score and iso-rate.

\subsubsection{Ablation studies}


We also conduct ablation studies to demonstrate the effectiveness of each part in our method. Specifically, we conduct individual experiments by removing different components, including before-training augmentation (BefAug), in-training augmentation (InAug), multi-frame detection module (MFD), and isolated-frame penalty (IFP).

From the multi-metric comparison in Table \ref{tab:freq1}, we observe that the IFP significantly reduces iso-rate while improving the correct detection score, as evident from the second and third rows. Additionally, the removal of MFD module, as indicated in the fourth row, results in a significant decrease in correct detection score, indicating its notable contribution, despite limited impact on iso-rate. Moreover, the fifth and sixth rows demonstrate that using data augmentation both before-training and in-training (BefAug \& InAug) leads to a significant improvement in performance, especially on more various augmented development set and test set.

\section{Conclusion}




In summary, we propose a basic RCNN-BLSTM backbone to classify each frame of audio. Then we enhance the stability and generalization ability of our base model through diverse data augmentation techniques. We also innovatively design a multi-frame detection (MFD) module and isolated-frame penalty (IFP) loss to improve the representation of local features and constrain outliers, resulting in improved model performance. These three aspects are crucial for our success in manipulated region detection challenge. In the future, to further improve the detection capabilities of our system, we will focus on enhancing the model's robustness against noise.

\bibliography{reference}

\begin{thebibliography}{27}
\expandafter\ifx\csname natexlab\endcsname\relax\def\natexlab#1{#1}\fi
\providecommand{\url}[1]{\texttt{#1}}
\providecommand{\href}[2]{#2}
\providecommand{\path}[1]{#1}
\providecommand{\DOIprefix}{doi:}
\providecommand{\ArXivprefix}{arXiv:}
\providecommand{\URLprefix}{URL: }
\providecommand{\Pubmedprefix}{pmid:}
\providecommand{\doi}[1]{\href{http://dx.doi.org/#1}{\path{#1}}}
\providecommand{\Pubmed}[1]{\href{pmid:#1}{\path{#1}}}
\providecommand{\bibinfo}[2]{#2}
\ifx\xfnm\relax \def\xfnm[#1]{\unskip,\space#1}\fi
\bibitem[{Wang et~al.(2017)Wang, Skerry-Ryan, Stanton, Wu, Weiss, Jaitly, Yang,
  Xiao, Chen, Bengio et~al.}]{wang2017tacotron}
\bibinfo{author}{Y.~Wang}, \bibinfo{author}{R.~Skerry-Ryan},
  \bibinfo{author}{D.~Stanton}, \bibinfo{author}{Y.~Wu}, \bibinfo{author}{R.~J.
  Weiss}, \bibinfo{author}{N.~Jaitly}, \bibinfo{author}{Z.~Yang},
  \bibinfo{author}{Y.~Xiao}, \bibinfo{author}{Z.~Chen},
  \bibinfo{author}{S.~Bengio}, et~al.,
\newblock \bibinfo{title}{Tacotron: Towards end-to-end speech synthesis},
\newblock \bibinfo{journal}{arXiv preprint arXiv:1703.10135}
  (\bibinfo{year}{2017}).
\bibitem[{Ren et~al.(2020)Ren, Hu, Tan, Qin, Zhao, Zhao, and
  Liu}]{ren2020fastspeech}
\bibinfo{author}{Y.~Ren}, \bibinfo{author}{C.~Hu}, \bibinfo{author}{X.~Tan},
  \bibinfo{author}{T.~Qin}, \bibinfo{author}{S.~Zhao},
  \bibinfo{author}{Z.~Zhao}, \bibinfo{author}{T.-Y. Liu},
\newblock \bibinfo{title}{Fastspeech 2: Fast and high-quality end-to-end text
  to speech},
\newblock \bibinfo{journal}{arXiv preprint arXiv:2006.04558}
  (\bibinfo{year}{2020}).
\bibitem[{Tan et~al.(2022)Tan, Chen, Liu, Cong, Zhang, Liu, Wang, Leng, Yi, He
  et~al.}]{tan2022naturalspeech}
\bibinfo{author}{X.~Tan}, \bibinfo{author}{J.~Chen}, \bibinfo{author}{H.~Liu},
  \bibinfo{author}{J.~Cong}, \bibinfo{author}{C.~Zhang},
  \bibinfo{author}{Y.~Liu}, \bibinfo{author}{X.~Wang},
  \bibinfo{author}{Y.~Leng}, \bibinfo{author}{Y.~Yi}, \bibinfo{author}{L.~He},
  et~al.,
\newblock \bibinfo{title}{Naturalspeech: End-to-end text to speech synthesis
  with human-level quality},
\newblock \bibinfo{journal}{arXiv preprint arXiv:2205.04421}
  (\bibinfo{year}{2022}).
\bibitem[{Shen et~al.(2023)Shen, Ju, Tan, Liu, Leng, He, Qin, Zhao, and
  Bian}]{shen2023naturalspeech}
\bibinfo{author}{K.~Shen}, \bibinfo{author}{Z.~Ju}, \bibinfo{author}{X.~Tan},
  \bibinfo{author}{Y.~Liu}, \bibinfo{author}{Y.~Leng}, \bibinfo{author}{L.~He},
  \bibinfo{author}{T.~Qin}, \bibinfo{author}{S.~Zhao},
  \bibinfo{author}{J.~Bian},
\newblock \bibinfo{title}{Naturalspeech 2: Latent diffusion models are natural
  and zero-shot speech and singing synthesizers},
\newblock \bibinfo{journal}{arXiv preprint arXiv:2304.09116}
  (\bibinfo{year}{2023}).
\bibitem[{Luong and Yamagishi(2020)}]{luong2020nautilus}
\bibinfo{author}{H.-T. Luong}, \bibinfo{author}{J.~Yamagishi},
\newblock \bibinfo{title}{Nautilus: a versatile voice cloning system},
\newblock \bibinfo{journal}{IEEE/ACM Transactions on Audio, Speech, and
  Language Processing} \bibinfo{volume}{28} (\bibinfo{year}{2020})
  \bibinfo{pages}{2967--2981}.
\bibitem[{Ma et~al.(2020)Ma, Su, Wang, and Lu}]{ma2020fpets}
\bibinfo{author}{D.~Ma}, \bibinfo{author}{Z.~Su}, \bibinfo{author}{W.~Wang},
  \bibinfo{author}{Y.~Lu},
\newblock \bibinfo{title}{Fpets: fully parallel end-to-end text-to-speech
  system},
\newblock in: \bibinfo{booktitle}{Proceedings of the AAAI Conference on
  Artificial Intelligence}, volume~\bibinfo{volume}{34}, \bibinfo{year}{2020},
  pp. \bibinfo{pages}{8457--8463}.
\bibitem[{Casanova et~al.(2022)Casanova, Weber, Shulby, Junior, G{\"o}lge, and
  Ponti}]{casanova2022yourtts}
\bibinfo{author}{E.~Casanova}, \bibinfo{author}{J.~Weber},
  \bibinfo{author}{C.~D. Shulby}, \bibinfo{author}{A.~C. Junior},
  \bibinfo{author}{E.~G{\"o}lge}, \bibinfo{author}{M.~A. Ponti},
\newblock \bibinfo{title}{Yourtts: Towards zero-shot multi-speaker tts and
  zero-shot voice conversion for everyone},
\newblock in: \bibinfo{booktitle}{International Conference on Machine
  Learning}, \bibinfo{organization}{PMLR}, \bibinfo{year}{2022}, pp.
  \bibinfo{pages}{2709--2720}.
\bibitem[{Li et~al.(2023)Li, Tu, and Xiao}]{li2023freevc}
\bibinfo{author}{J.~Li}, \bibinfo{author}{W.~Tu}, \bibinfo{author}{L.~Xiao},
\newblock \bibinfo{title}{Freevc: Towards high-quality text-free one-shot voice
  conversion},
\newblock in: \bibinfo{booktitle}{ICASSP 2023-2023 IEEE International
  Conference on Acoustics, Speech and Signal Processing (ICASSP)},
  \bibinfo{organization}{IEEE}, \bibinfo{year}{2023}, pp.
  \bibinfo{pages}{1--5}.
\bibitem[{Wang et~al.(2021)Wang, Li, Tang, Yin, Wan, Yu, and
  Ma}]{wang2021towards}
\bibinfo{author}{C.~Wang}, \bibinfo{author}{Z.~Li}, \bibinfo{author}{B.~Tang},
  \bibinfo{author}{X.~Yin}, \bibinfo{author}{Y.~Wan}, \bibinfo{author}{Y.~Yu},
  \bibinfo{author}{Z.~Ma},
\newblock \bibinfo{title}{Towards high-fidelity singing voice conversion with
  acoustic reference and contrastive predictive coding},
\newblock \bibinfo{journal}{arXiv preprint arXiv:2110.04754}
  (\bibinfo{year}{2021}).
\bibitem[{Wu et~al.(2020)Wu, Liu, and Lee}]{wu2020defense}
\bibinfo{author}{H.~Wu}, \bibinfo{author}{A.~T. Liu}, \bibinfo{author}{H.-y.
  Lee},
\newblock \bibinfo{title}{Defense for black-box attacks on anti-spoofing models
  by self-supervised learning},
\newblock \bibinfo{journal}{arXiv preprint arXiv:2006.03214}
  (\bibinfo{year}{2020}).
\bibitem[{Wang and Yamagishi(2021)}]{wang2021comparative}
\bibinfo{author}{X.~Wang}, \bibinfo{author}{J.~Yamagishi},
\newblock \bibinfo{title}{A comparative study on recent neural spoofing
  countermeasures for synthetic speech detection},
\newblock \bibinfo{journal}{arXiv preprint arXiv:2103.11326}
  (\bibinfo{year}{2021}).
\bibitem[{Peng et~al.(2021)Peng, Li, and Lee}]{peng2021pairing}
\bibinfo{author}{Z.~Peng}, \bibinfo{author}{X.~Li}, \bibinfo{author}{T.~Lee},
\newblock \bibinfo{title}{Pairing weak with strong: Twin models for defending
  against adversarial attack on speaker verification.},
\newblock in: \bibinfo{booktitle}{Interspeech}, \bibinfo{year}{2021}, pp.
  \bibinfo{pages}{4284--4288}.
\bibitem[{Yi et~al.(2023)Yi, Tao, Fu, Yan, Wang, Wang, Zhang, Zhang, Zhao, Ren,
  Xu, Zhou, Gu, Wen, Liang, Lian, and Li}]{deepfake}
\bibinfo{author}{J.~Yi}, \bibinfo{author}{J.~Tao}, \bibinfo{author}{R.~Fu},
  \bibinfo{author}{X.~Yan}, \bibinfo{author}{C.~Wang},
  \bibinfo{author}{T.~Wang}, \bibinfo{author}{C.~Y. Zhang},
  \bibinfo{author}{X.~Zhang}, \bibinfo{author}{Y.~Zhao},
  \bibinfo{author}{Y.~Ren}, \bibinfo{author}{L.~Xu}, \bibinfo{author}{J.~Zhou},
  \bibinfo{author}{H.~Gu}, \bibinfo{author}{Z.~Wen},
  \bibinfo{author}{S.~Liang}, \bibinfo{author}{Z.~Lian},
  \bibinfo{author}{H.~Li},
\newblock \bibinfo{title}{Add 2023: the second audio deepfake detection
  challenge, accepted by ijcai 2023 workshop on deepfake audio detection and
  analysis(dada2023)}  (\bibinfo{year}{2023}).
\bibitem[{Zeng et~al.(2022{\natexlab{a}})Zeng, Wang, Kong, Feng, Zhao, and
  Wang}]{zeng4051713deletion}
\bibinfo{author}{C.~Zeng}, \bibinfo{author}{Z.~Wang},
  \bibinfo{author}{S.~Kong}, \bibinfo{author}{S.~Feng},
  \bibinfo{author}{N.~Zhao}, \bibinfo{author}{J.~Wang},
\newblock \bibinfo{title}{Deletion and insertion tampering detection of digital
  audio based on enf fluctuating super vector},
\newblock \bibinfo{journal}{Available at SSRN 4051713}
  (\bibinfo{year}{2022}{\natexlab{a}}).
\bibitem[{Zeng et~al.(2022{\natexlab{b}})Zeng, Kong, Wang, Wan, and
  Chen}]{zeng2022}
\bibinfo{author}{C.~Zeng}, \bibinfo{author}{S.~Kong},
  \bibinfo{author}{Z.~Wang}, \bibinfo{author}{X.~Wan},
  \bibinfo{author}{Y.~Chen},
\newblock \bibinfo{title}{Digital audio tampering detection based on enf
  spatio-temporal features representation learning},
\newblock \bibinfo{journal}{arXiv preprint arXiv:2208.11920}
  (\bibinfo{year}{2022}{\natexlab{b}}).
\bibitem[{Gu et~al.(2018)Gu, Wang, Kuen, Ma, Shahroudy, Shuai, Liu, Wang, Wang,
  Cai et~al.}]{gu2018recent}
\bibinfo{author}{J.~Gu}, \bibinfo{author}{Z.~Wang}, \bibinfo{author}{J.~Kuen},
  \bibinfo{author}{L.~Ma}, \bibinfo{author}{A.~Shahroudy},
  \bibinfo{author}{B.~Shuai}, \bibinfo{author}{T.~Liu},
  \bibinfo{author}{X.~Wang}, \bibinfo{author}{G.~Wang},
  \bibinfo{author}{J.~Cai}, et~al.,
\newblock \bibinfo{title}{Recent advances in convolutional neural networks},
\newblock \bibinfo{journal}{Pattern recognition} \bibinfo{volume}{77}
  (\bibinfo{year}{2018}) \bibinfo{pages}{354--377}.
\bibitem[{Hochreiter and Schmidhuber(1997)}]{6795963}
\bibinfo{author}{S.~Hochreiter}, \bibinfo{author}{J.~Schmidhuber},
\newblock \bibinfo{title}{Long short-term memory},
\newblock \bibinfo{journal}{Neural Computation} \bibinfo{volume}{9}
  (\bibinfo{year}{1997}) \bibinfo{pages}{1735--1780}.
  \DOIprefix\doi{10.1162/neco.1997.9.8.1735}.
\bibitem[{Wang et~al.(2022)Wang, Yang, Zeng, Kong, Feng, and Zhao}]{wang2022}
\bibinfo{author}{Z.~Wang}, \bibinfo{author}{Y.~Yang},
  \bibinfo{author}{C.~Zeng}, \bibinfo{author}{S.~Kong},
  \bibinfo{author}{S.~Feng}, \bibinfo{author}{N.~Zhao},
\newblock \bibinfo{title}{Shallow and deep feature fusion for digital audio
  tampering detection},
\newblock \bibinfo{journal}{EURASIP Journal on Advances in Signal Processing}
  \bibinfo{volume}{2022} (\bibinfo{year}{2022}) \bibinfo{pages}{69}.
\bibitem[{Winursito et~al.(2018)Winursito, Hidayat, and
  Bejo}]{winursito2018improvement}
\bibinfo{author}{A.~Winursito}, \bibinfo{author}{R.~Hidayat},
  \bibinfo{author}{A.~Bejo},
\newblock \bibinfo{title}{Improvement of mfcc feature extraction accuracy using
  pca in indonesian speech recognition},
\newblock in: \bibinfo{booktitle}{2018 International Conference on Information
  and Communications Technology (ICOIACT)}, \bibinfo{organization}{IEEE},
  \bibinfo{year}{2018}, pp. \bibinfo{pages}{379--383}.
\bibitem[{Hamza et~al.(2022)Hamza, Javed, Iqbal, Kryvinska, Almadhor, Jalil,
  and Borghol}]{hamza2022}
\bibinfo{author}{A.~Hamza}, \bibinfo{author}{A.~R.~R. Javed},
  \bibinfo{author}{F.~Iqbal}, \bibinfo{author}{N.~Kryvinska},
  \bibinfo{author}{A.~S. Almadhor}, \bibinfo{author}{Z.~Jalil},
  \bibinfo{author}{R.~Borghol},
\newblock \bibinfo{title}{Deepfake audio detection via mfcc features using
  machine learning},
\newblock \bibinfo{journal}{IEEE Access} \bibinfo{volume}{10}
  (\bibinfo{year}{2022}) \bibinfo{pages}{134018--134028}.
\bibitem[{Tak et~al.(2021)Tak, Patino, Todisco, Nautsch, Evans, and
  Larcher}]{tak2021end}
\bibinfo{author}{H.~Tak}, \bibinfo{author}{J.~Patino},
  \bibinfo{author}{M.~Todisco}, \bibinfo{author}{A.~Nautsch},
  \bibinfo{author}{N.~Evans}, \bibinfo{author}{A.~Larcher},
\newblock \bibinfo{title}{End-to-end anti-spoofing with rawnet2},
\newblock in: \bibinfo{booktitle}{ICASSP 2021-2021 IEEE International
  Conference on Acoustics, Speech and Signal Processing (ICASSP)},
  \bibinfo{organization}{IEEE}, \bibinfo{year}{2021}, pp.
  \bibinfo{pages}{6369--6373}.
\bibitem[{Ustubioglu et~al.(2023)Ustubioglu, Tahaoglu, and
  Ulutas}]{ustubioglu2023}
\bibinfo{author}{B.~Ustubioglu}, \bibinfo{author}{G.~Tahaoglu},
  \bibinfo{author}{G.~Ulutas},
\newblock \bibinfo{title}{Detection of audio copy-move-forgery with novel
  feature matching on mel spectrogram},
\newblock \bibinfo{journal}{Expert Systems with Applications}
  \bibinfo{volume}{213} (\bibinfo{year}{2023}) \bibinfo{pages}{118963}.
\bibitem[{Yang et~al.(2023)Yang, Liu, and Cao}]{yang2023}
\bibinfo{author}{D.~Yang}, \bibinfo{author}{M.~Liu}, \bibinfo{author}{M.~Cao},
\newblock \bibinfo{title}{Fast blind audio copy-move detection and localization
  using local feature tensors in noise},
\newblock \bibinfo{journal}{arXiv preprint arXiv:2302.07584}
  (\bibinfo{year}{2023}).
\bibitem[{Wu et~al.(2022)Wu, Kuo, Zheng, Hung, Lee, Tsao, Wang, and
  Meng}]{wu2022partially}
\bibinfo{author}{H.~Wu}, \bibinfo{author}{H.-C. Kuo},
  \bibinfo{author}{N.~Zheng}, \bibinfo{author}{K.-H. Hung},
  \bibinfo{author}{H.-Y. Lee}, \bibinfo{author}{Y.~Tsao},
  \bibinfo{author}{H.-M. Wang}, \bibinfo{author}{H.~Meng},
\newblock \bibinfo{title}{Partially fake audio detection by
  self-attention-based fake span discovery},
\newblock in: \bibinfo{booktitle}{ICASSP 2022-2022 IEEE International
  Conference on Acoustics, Speech and Signal Processing (ICASSP)},
  \bibinfo{organization}{IEEE}, \bibinfo{year}{2022}, pp.
  \bibinfo{pages}{9236--9240}.
\bibitem[{Skerry-Ryan et~al.(2018)Skerry-Ryan, Battenberg, Xiao, Wang, Stanton,
  Shor, Weiss, Clark, and Saurous}]{skerryryan2018referenceencoder}
\bibinfo{author}{R.~Skerry-Ryan}, \bibinfo{author}{E.~Battenberg},
  \bibinfo{author}{Y.~Xiao}, \bibinfo{author}{Y.~Wang},
  \bibinfo{author}{D.~Stanton}, \bibinfo{author}{J.~Shor},
  \bibinfo{author}{R.~J. Weiss}, \bibinfo{author}{R.~Clark},
  \bibinfo{author}{R.~A. Saurous}, \bibinfo{title}{Towards end-to-end prosody
  transfer for expressive speech synthesis with tacotron},
  \bibinfo{year}{2018}. \href{http://arxiv.org/abs/1803.09047}{{\tt
  arXiv:1803.09047}}.
\bibitem[{Snyder et~al.(2015)Snyder, Chen, and Povey}]{musan2015}
\bibinfo{author}{D.~Snyder}, \bibinfo{author}{G.~Chen},
  \bibinfo{author}{D.~Povey}, \bibinfo{title}{{MUSAN}: {A} {M}usic, {S}peech,
  and {N}oise {C}orpus}, \bibinfo{year}{2015}.
  \href{http://arxiv.org/abs/1510.08484}{{\tt arXiv:1510.08484}},
  \bibinfo{note}{arXiv:1510.08484v1}.
\bibitem[{Baevski et~al.(2020)Baevski, Zhou, Mohamed, and
  Auli}]{baevski2020wav2vec}
\bibinfo{author}{A.~Baevski}, \bibinfo{author}{H.~Zhou},
  \bibinfo{author}{A.~Mohamed}, \bibinfo{author}{M.~Auli},
  \bibinfo{title}{wav2vec 2.0: A framework for self-supervised learning of
  speech representations}, \bibinfo{year}{2020}.
  \href{http://arxiv.org/abs/2006.11477}{{\tt arXiv:2006.11477}}.

\end{thebibliography}
\bibliographystyle{plain}
\renewcommand{\bibfont}{\tiny}
\end{document}